\documentclass[a4paper,11pt]{article}

\usepackage{pos}
\usepackage{ulem}

\title{Reflections on an  M-theoretic Emergence Proposal} 

\author*[a,b]{Ralph Blumenhagen}
\author[a]{Niccol\`o Cribiori}
\author[a,b]{Aleksandar Gligovic}
\author[a,c]{Antonia Paraskevopoulou}

\affiliation[a]{Max-Planck-Institut f\"ur Physik,\\
Boltzmannstr. 8,  85748 Garching, Germany }
\affiliation[b]{Exzellenzcluster ORIGINS, \\ Boltzmannstr. 2, D-85748 Garching, Germany}
\affiliation[c]{Fakult{\"a}t f{\"u}r Physik, Ludwig-Maximilians-Universit{\"a}t M\"unchen, \\ 
Theresienstr.~37, 80333 M\"unchen, Germany}

\emailAdd{blumenha@mpp.mpg.de}
\emailAdd{cribiori@mpp.mpg.de}
\emailAdd{aglig@mpp.mpg.de}
\emailAdd{aparaske@mpp.mpg.de}

\abstract{
In a pedagogical manner, we review recent developments in the investigation of the Emergence Proposal. 
Although it is fair to say that this idea is still at an exploratory level and a fully coherent picture has yet to be developed, we put it into perspective to previous work on the swampland program  and on emergence in QG. 
In view of the emergent string conjecture, we argue and provide evidence that it is not the emergent string but rather the decompactification limit which is a natural candidate for the potential realization of the Emergence Proposal. 
This resonates in a compelling way with old ideas of emergence in M(atrix) theory and gives rise to a number of further speculations.}

\FullConference{Corfu Summer Institute 2023 "School and Workshops on Elementary Particle Physics and Gravity" (CORFU2023)\\
 23 April - 6 May , and 27 August - 1 October, 2023\\
Corfu, Greece\\}

\newcommand{\eq}[1]{\begin{equation}
	\begin{split} #1 \end{split}
	\end{equation}}

\newcommand{\ov}{\overline}

\begin{document}

\renewcommand{\hookAfterAbstract}{
    \par\bigskip\bigskip\bigskip
    Report number: MPP-2024-73
}
\maketitle

\section{Introduction}
\label{sec_intro}

That the notion of emergence may be relevant in theories of Quantum Gravity (QG) is not a new idea.  
While different interpretations of emergence might be encountered across different areas of study, in the context of QG it has been defined \cite{Butterfield:2011,Oriti:2018dsg} as the appearance of properties of a system that are novel with respect to other (more fundamental) descriptions of the same system and robust in the sense of characterizability and reproducibility.
However, there remains the more philosophical question of what is actually meant by emergence, if it  really describes a novel phenomenon or if it is just a relative, rather epistemological notion reflecting the state of ignorance about all implications of a given theory.
Not entering this discussion, we take a more pragmatic point of view and consider a standard string theory example to elaborate on how emergence should be understood in the course of this article.

An open string stretched between two $D$-branes has a tower of excitations whose massless modes are gauge fields confined to the branes.
Computing the one-loop annulus diagram and applying loop channel - tree channel equivalence, the amplitude encodes the tree-level exchange of gravitons, i.e.~of closed string excitations.
Here, we say that QG is emerging via quantum effects from another theory not having the graviton as fundamental degree of freedom. 
On the other side, in weakly coupled perturbative string theory one usually concludes that, due to loop channel - tree channel equivalence again, a theory of just open strings is not consistent and one always needs to include the closed strings, as well.
This is confirmed by all existing superstring theories in 10D and their compactifications. 
Then, the leading order gravitational interaction is just a tree-level effect in closed string theory.
Since the quantization of closed strings includes gravity, in these theories we  say that QG is not truly emerging.\footnote{One could say that gravity emerges  from the quantization of closed strings, but this is not how we want to understand it.}

A lesson we draw from this is that to realize emergence in the above sense we need a theory featuring $D$-branes as fundamental light degrees of freedom,  with closed strings being heavy.
As mentioned, this cannot happen in weakly coupled string theory, $g_s=e^\phi\ll 1$, so that one is guided towards other asymptotic limits of QG, that might be potentially realizing such a hierarchical pattern of mass scales.
Thanks to recent progress in the swampland program (see e.g.~\cite{Palti:2019pca,vanBeest:2021lhn,Agmon:2022thq} for reviews), we now have a more systematic understanding of such infinite distance limits. 
There are a couple of well established swampland conjectures particularly dealing with them, known  as the swampland distance conjecture \cite{Ooguri:2006in} and one of its refinements, the emergent string conjecture \cite{Lee:2019wij}.

As will be reviewed in section \ref{sec_two}, the lesson one can draw from them is that in infinite distance limits, with a parameter $t\gg 1$, the degrees of freedom of QG will show a hierarchical pattern such that one can distinguish between light and heavy modes, with the former being the fundamental quantum degrees of freedom while the latter can be thought of as classical.
In terms of the naturally small parameter $g=1/t\ll 1$, the mass scale of these modes behave as
\begin{equation}
\label{pertandclass}
m_{\rm pert}\simeq g^\alpha \Lambda \,, \qquad m_{\rm class}\simeq\frac{\Lambda}{g^\beta}
\end{equation}
with $\alpha\ge 0$, $\beta>0$ and $\Lambda$ a characteristic mass scale of the limit. 
This is the behavior also known from perturbative quantum field theories (QFTs), where we distinguish between the classical non-perturbative contributions to the path integral and the light quantum fluctuations around them.
Hence, it is in these limits that we can expect to find something like a perturbative QG theory that shows certain resemblance to quantum field theory.

From this perspective, the usual perturbative string theory is just the perturbative QG theory that arises in the small string coupling regime, $g_s=e^\phi\ll 1$, i.e.~the infinite distance limit $\phi\to-\infty$ for the dilaton. The characteristic mass scale $\Lambda$ is given by the string scale $\Lambda=M_s=(\alpha')^{-1/2}$.
The perturbative states are the vibration modes of the string and upon compactification also its Kaluza-Klein (KK) and winding modes.
The non-perturbative states are the various $p$-branes with tension $T_p\simeq M_s^{p+1}/g_s^\beta$, $\beta=1,2$. 
For $D$-branes with $\beta=1$, the quantum fluctuations (of the closed string modes) around them can be described by open string modes and in fact the $D$-brane can be characterized as a coherent state (boundary state in CFT) of closed strings. 
It is important to notice that the vibration modes of the $p$-branes themselves are of the scale of the brane tension and hence are heavy and freeze out in the limit $g_s\to 0$.
While they seem similar at first sight, this example reveals  that QG is differing from QFT in one important aspect, namely that it does not feature a finite number of perturbative modes but rather infinite towers thereof already at first quantization.

To connect to the notion of emergence described initially, the crucial question is whether there exist asymptotic limits of QG where the light states are not string excitations. 
What is the value of $\Lambda$ in these cases? 
If such non-string limits exist, then the highly non-trivial question of how one can mathematically describe these new perturbative QG theories arises. 
With no string loop expansion available, as well as the techniques developed so far for it evaluation available, how can one determine e.g.~the low-energy effective action for the massless modes?
As we will describe, the answer to these questions seems to be closely related to
the two aforementioned swampland conjectures, the notion of species scale \cite{Dvali:2007hz,Dvali:2007wp} and the so-called Emergence Proposal \cite{Heidenreich:2017sim,Grimm:2018ohb,Heidenreich:2018kpg}.
The latter suggests that the kinetic terms in the low-energy effective action of (all) QG theories are emerging quantum mechanically from integrating out states below a certain ultraviolet (UV) scale.

It is fair to say that the Emergence Proposal is currently on less firm ground than other swampland conjectures.
It is the purpose of this article to review and conceptually reflect on some recent advances \cite{Blumenhagen:2023tev,Blumenhagen:2023xmk,Blumenhagen:2024ydy} in concretizing this general idea and to connect it to the concept of emergence mentioned at the very beginning of this introduction. 
At the end of section \ref{sec:emMth}, we will also comment on the recent work \cite{Hattab:2023moj,Hattab:2024thi} employing an approach slightly different form ours.
Admittedly, part of the results of \cite{Blumenhagen:2023tev,Blumenhagen:2023xmk,Blumenhagen:2024ydy} build upon previous seminal work by e.g.~Green-Gutperle-Vanhove \cite{Green:1997as} (and then \cite{Green:1997di,Kiritsis:1997em,Russo:1997mk,Pioline:1997pu,Obers:1998fb,Obers:1999um,Angelantonj:2011br}) and Gopakumar-Vafa
\cite{Gopakumar:1998ii,Gopakumar:1998jq}, which however at that time did not explicitly emphasize the relation to emergence, let alone the swampland program.

\section{Preliminaries}
\label{sec_two}

In this section we first lay out some of the basic notions from the swampland program predating the formulation of the Emergence Proposal.
Then, we discuss the Emergence Proposal in its initial form and argue that a refined M-theoretic version of it has a real chance of being realized.

\subsection{The swampland distance conjecture}

The swampland distance conjecture \cite{Ooguri:2006in} states that when approaching points at infinite distance in the  moduli space of an effective field theory (EFT) arising from a viable QG theory, an infinite tower of states becomes exponentially light. Denoting with $M_n$ the mass of the $n$-th level of the tower, the statement is that for $\phi\to \infty$ we have
\begin{equation}
M_{n} \simeq f(n)\, e^{-\alpha \phi}\,,
\end{equation}
where $\alpha$ is of order one in natural units where $M_{\rm pl}=1$.
Since the EFT comes with an UV cut-off above which all the states have been integrated out, this implies a breakdown of the effective description for field excursions $\phi\gtrsim\alpha^{-1}$. 
Hence, an EFT derived from QG only has a finite range of validity.

In string theory two different types of such towers are usually encountered, one is genuinely stringy in nature, while the other has to do with the fact that string theory is a higher dimensional theory. 
Consider for instance critical superstring theory, for which the relation between the 10D Planck scale and the string scale is
\begin{equation}
\label{MstoMp10d}
M_{\rm s}\simeq M_{\rm pl} \,g_s^{1/4}\simeq M_{\rm pl}\, e^{\phi/4} \,.
\end{equation}
Therefore, in the infinite distance limit of weak string coupling, $\phi\to -\infty$, (for fixed Planck scale) the tower of string excitations of mass $M_n=\sqrt{n} M_s$ becomes exponentially light. 
Asymptotically this tower has an exponential degeneracy of states at mass level $n$ given by ${\rm deg}_n\simeq\exp(\sqrt{n})$.

Compactifying string theory to lower dimensions, like e.g.~to 9D on a circle, the sizes of the compact dimensions will appear as extra scalars in the EFT. Taking one of these scalars to infinite distance leads to a decompactification limit, for which the KK modes become exponentially light. 
To see this, consider the circle compactification on a radius of size $R$. Setting $\phi=\gamma \log(M_{\rm pl} R)$, where $\gamma$ is fixed by canonically normalizing the kinetic term,  the mass of the KK tower is given by
\begin{equation}
M_n\simeq\frac{n}{R} \simeq M_{\rm pl}\, n\, e^{-\phi/\gamma} \,.
\end{equation}
Hence, the KK tower shows an exponential scaling behavior with a dependence on the level $n$ and a polynomial degeneracy ${\rm deg}_n$ (here ${\rm deg}_n=1$ for a single $S^1$).

\subsection{The Emergent String Conjecture}

From these two simple string theory examples, it might seem as a big leap to conjecture that this is  essentially already the exhaustive list of different behaviors in any theory of QG. Nevertheless, this is precisely what the emergent string conjecture \cite{Lee:2019wij} states:

\begin{quotation}
\noindent
{\it Any infinite distance limit in QG is either an emerging string limit, where a fundamental string tower accompanied by particle-like towers becomes light, or a decompactification limit, where the lightest tower shows the behavior of a KK tower.}
\end{quotation}

\noindent 
Evidence for this conjecture was initially collected for Calabi-Yau  compactifications of M-theory and type IIA superstring theory in infinite distance limits in their vector-multiplet moduli space. 
To appreciate the meaning of this conjecture, it is important to notice that these two limiting behaviors can come in various disguises. 

\paragraph*{Emergent string limit}
The most obvious such limit is the aforementioned coscaled weak coupling limit
\begin{equation}
g_s\to \lambda g_s\,,\qquad M_s\to \lambda^{1/4} M_s\,,
\end{equation}
of type IIA/IIB string theory in 10D.  Here we have $\lambda\ll 1$ while keeping  the 10D Planck scale constant. 
For the type IIB superstring, one can also consider the analogous strong coupling regime, $\lambda\gg 1$, leading to an emergent string limit where the string tower comes from $D1$-branes \cite{Blumenhagen:2023xmk}. 
This can be generalized to compactifications on $k$-dimensional tori down to $d=10-k$ dimensions.
In this case, the coscaled emergent $D1$-string limit is
\begin{equation}
g_s\to \lambda g_s\,,\qquad  M_s\to \lambda^{\frac{d-6}{ 2(d-2)}} M_s\,, \qquad \rho_1\to \lambda^{\frac 12} \rho_1 \,,\qquad \rho_i\to \lambda^{\frac 12} \rho_i\,,
\end{equation}
where $\rho_i$ denote the internal radii in string units.
After applying the map
\begin{equation}
\label{map2bm}
M_s^2=M_*^2\, r_{11}\,,\qquad g_s=\frac{r_{11}}{r_1}\,,\qquad   \rho_1=\frac{1}{ r_1 \,r^{1/2}_{11}}\,,\qquad
\rho_i=r_i \,r^{1/2}_{11}\,,
\end{equation}  
between M-theory (on $T^2\times T^{k-1}$) and type IIB (on $T^{k}$), one can transform this scaling to the M-theory quantities
\eq{
      r_{11}\to \lambda^{\frac{1}{3}} r_{11}\,,\qquad M_*\to \lambda^{\frac{(d-8)}{3(d-2)}} M_*\,,\qquad  r_1\to \lambda^{-{\frac{2}{3}}} r_1\,,\qquad r_i \to \lambda^{\frac{1}{3}} r_i\,,
 }     
where $r_{11}$, $r_1$ and the $r_i$  denote the internal radii in 11D Planck units and  $M_*$ the 11D Planck scale.  In this  limit,  the lightest states are the $D1$-branes, whose mass scale is
 \eq{
   M_{D1}= T_{D1}^{1/2}\simeq {\frac{M_s}{g_s^{1/2}}}\simeq {\frac{M_{\rm pl}^{(d)}}{\lambda^{\frac{2}{d-2}}}}\,.
     }

However, there also exist more involved limits of this type.
Consider e.g. the compactification of the type IIA superstring on a Calabi-Yau which is $K3$-fibered over a $\mathbb P^1$.
Let us denote the size of the base as $t_b$ so that the total volume of the Calabi-Yau is given by $V=t_b \tau_{K3}+\ldots$ where $\tau_{K3}$ indicates the size of the $K3$ fiber. 
Now, we can consider the infinite distance limit where we scale $t_b\to \lambda t_b$ with $\lambda\gg 1$ while keeping $\tau_{K3}$ finite. 
To maintain also the 4D Planck scale finite, $M_{\rm pl}^2=M_s^2 V/g_s^2$, we have to coscale $g_s\to \lambda^{1/2} g_s$. 
This means that we are taking a  limit with a large string coupling.
Scanning through the list of states, we see that the lightest modes scale like $M_s\simeq M_{\rm pl}/\lambda^{1/2}$ and are given by  the excitations of a string resulting from wrapping the $NS5$-brane on $K3$, together with the particle-like states of $D0$- and transverse (to the large $\mathbb P^1$) $D2$- and $D4$-branes. 
The excitations of the type IIA fundamental string scale like $M\simeq\lambda^0$ and are parametrically heavier. 
One can show that in this limit there exists a weakly coupled heterotic dual model compactified on $K3\times T^2$ such that the complexified K\"ahler modulus, $T_B=t_B+i b$, is mapped to the dilaton of the heterotic string, $S=\exp(-\phi_H)+i B$. 
The light towers of states are mapped to the heterotic string excitations and to the KK and winding modes on $K3\times T^2$. 
Hence this type IIA infinite distance limit is an emergent string limit, where a (dual) fundamental string is among the lightest modes.

\paragraph*{Decompactification limit}

The type IIA K\"ahler moduli space of Calabi-Yau compactifications also admits a
coscaled decompactification limit, which is given by scaling all of its K\"ahler moduli isotropically as $t_I\to \lambda^{2/3} t_I$ and coscaling the dilaton as $g_s\to \lambda g_s$ to keep the 4D Planck scale finite. 
For later purposes let us discuss this strong coupling limit more generally, i.e.~upon compactifying type IIA string theory on an internal manifold
$X$ of dimension $k$.  Since  this limit will be related to M-theory,
we recall the dictionary between the strong coupling limit of the type IIA superstring and M-theory.
The string scale $M_s$ and coupling $g_s$ are given in terms of the 11D Planck scale $M_*$ and of the size $r_{11}$ of the eleventh direction as
\begin{equation}
M_s^2=M_*^2\, r_{11}\,,\qquad g_s=r_{11}^{\frac{3}{2}}\,.
\end{equation}
Let us consider the  strong coupling limit, $\lambda\to\infty$, such that the $d=10-k$ dimensional Planck scale $M_{\rm pl}^{(d)}$ and the size of the internal space remain finite in units of $M_*$.
In terms of the type IIA quantities it reads
\begin{equation}
  g_s\to \lambda  g_s\,,\qquad M_s\to \lambda^{\frac{d-4}{3(d-2)}} \,M_s\,,\qquad
  \rho_I\to \lambda^{\frac{1}{3}} \rho_I\,.
\end{equation}  
Note that $d=4$ is special in the sense that the string scale does not scale with $\lambda$.
This can be translated to the  M-theory quantities as
\begin{equation}
r_{11}\to \lambda^{\frac{2}{3}}  r_{11}\,,\qquad M_*\to \frac{M_*}{ \lambda^{\frac{2}{3(d-2)}}}\,,\qquad r_I\to  r_I\,,
\end{equation}
where $r_I$ denote the  radii of the internal  space $X$ in units of $M_*$.
This means that all length scales of $X$ are scaled isotropically.
One can show that the $(d+1)$-dimensional Planck scale $M_{\rm pl}^{(d+1)}$
scales in the same way as $M_*$.

From the M-theory perspective this particular type IIA strong coupling limit corresponds to decompactification from $d$ to $d+1$ dimensions. 
The lightest tower of states are particle-like  $D0$-branes, or equivalently KK states of the eleventh direction of mass
\eq{
  M_{D0}\simeq {\frac{M_s}{g_s}} \simeq {\frac{M_{\rm pl}^{(d)}}{\lambda^{\frac{2(d-1)}{3(d-2)}}}}\,.
}  
The next lightest states are arising from  wrapped $D2$- and $NS5$-branes, having a mass scale  $M_{D2,NS5}\simeq M_s/g_s^{1/3} \simeq  M_{\rm pl}^{(d)}/\lambda^{\frac{2}{3(d-2)}}\simeq M_*$. 
KK modes along other compact directions $I$ also have mass $M_{\rm KK}\simeq M_*/r_I$.
All other states, like wrapped $D4$-branes or the fundamental string, are parametrically heavier. Hence, this is a typical example of a decompactification limit where the lightest tower is (dual) to a KK tower of particle-like states.

\vspace{0.1cm}
The evidence for the emergent string conjecture comes mostly
from string theory examples.\footnote{See \cite{Basile:2023blg} for a recent attempt to recover the emergent string conjecture via bottom-up arguments based on black hole thermodynamics.}
It is nevertheless remarkable that this string lamppost approach already led to more than just emergent string limits, namely the existence of decompactification limits. 
As we have seen, one of them is closely related to the M-theory corner of the known string duality
diagram. 
It would be interesting to know how generic this limit is. The suspicion is that all infinite distance decompactification limits are combinations of the M-theory limit and its conventional further decompactification limits of additional compact directions.

\subsection{The species scale}
\label{sec_species}

Naively one would think that the mass scale where QG effects become important is the Planck mass. 
However, already the weakly coupled string theory shows the relevance of another mass scale, namely the string scale $M_s$, which in 10D is related to the Planck scale via \eqref{MstoMp10d}. 
For fixed Planck scale and in the infinite distance limit $g_s\to 0$,  the string scale can become arbitrarily small. 
This means that new, QG related physics will already occur at an energy scale well below the Planck scale.

One can ask the question whether also the decompactification limit comes with such an intermediate  mass scale and how it can be determined. 
Quite intriguingly, the general appearance of an effective QG cut-off in the case of a large number of light states was pointed out independently of any string theory reasoning in \cite{Dvali:2007hz,Dvali:2007wp} (see also \cite{Veneziano:2001ah} for earlier work).
When considering the quantum corrections to the 4D graviton propagator due to the coupling of a large number $N_{\rm sp}$ of light species to gravity, the combination $N_{\rm sp} p^2/M_{\rm pl}^2$ appears, with $p$ the momentum involved in the process.
If this combination becomes of order one  perturbation theory definitely breaks down revealing the mass scale $\tilde{\Lambda}\simeq M_{\rm pl}/\sqrt{N_{\rm sp}}<M_{\rm pl}$
where quantum effects of gravity become important. This is the
so-called species scale, which in $d$ dimensions reads
\begin{equation}
\label{speciesscale}
\tilde{\Lambda}\simeq \frac{M_{\rm pl}}{N_{\rm sp}^{\frac{1}{d-2}}}\,.
\end{equation}
For a given tower, the number of light species with mass below the species scale can effectively be counted as
\begin{equation}
\label{speciesnumber}
N_{\rm sp}=\#( m\le \tilde\Lambda )\,.
\end{equation}
Then, the latter two equations can be solved for the two unknowns $N_{\rm sp}$ and $\tilde\Lambda$. 
For KK towers this definition indeed gives the correct species scale, whereas for string towers it gives an extra multiplicative $\log$-factor that is not expected to be physical, for reasons that we will review in the following.

It turns out that one could alternatively think of  the species scale as the scale where  quantum corrections to the leading order Einstein-Hilbert term become relevant. 
One way of seeing this is to define the species scale as the radius $r_0=1/\tilde\Lambda$ of the minimal-sized black hole that can be described within the EFT. 
The mass and Bekenstein-Hawking entropy of such a black hole are
\begin{equation}
\label{BekensteinH}
M_{\rm BH}= \frac{M_{\rm pl}^{d-2}}{\tilde\Lambda^{d-3}}\,,\qquad\qquad  S_{\rm BH}=\frac {M_{\rm pl}^{d-2}}{\tilde\Lambda^{d-2}}\,.
\end{equation}
The number of species is defined via the statistical entropy as
\begin{equation}
\label{entropystat}
S_{\rm BH}=\log\Omega(M_{\rm BH}) =:N_{\rm sp}\,,
\end{equation}
where $\Omega(M_{\rm BH})$ is the number of ways the macroscopic
black hole of mass $M_{\rm BH}$ can be realized by the microstates.
Note that this definition of the number of species 
satisfies  the relation \eqref{speciesscale} and has also been developed into a more complete thermodynamic picture in \cite{Cribiori:2023ffn,Basile:2024dqq}.

Let us employ this definition for a string tower with mass levels $M=M_s \sqrt{N}$ and degeneracy ${\rm deg}_N$.
For sufficiently large mass levels $N$, one can use the asymptotic expansion
\begin{equation}
{\rm deg}_N\sim  e^{\beta \sqrt{N}}\,.
\end{equation}
Now, the excitation level required for the  black hole mass is
\begin{equation}
\label{oscinumberBH}
\sqrt{N_{\rm BH}}\simeq \frac{M_{\rm BH}}{M_s}\simeq \frac{M_{\rm pl}^{d-2}}{M_s \,\tilde\Lambda^{d-3}}\,,
\end{equation}
which for small $g_s$ is expected to be a very large number.
Then, up to a $\beta$ factor, the black hole  entropy is
\begin{equation}
S_{\rm BH}\simeq \log\left( {\rm deg}_{N_{\rm BH}} \right)\simeq    \sqrt{N_{\rm BH}} \,.
\end{equation}
Setting this equal to the Bekenstein-Hawking entropy \eqref{BekensteinH} and
using \eqref{oscinumberBH} gives at leading order 
\begin{equation}
\tilde\Lambda\simeq M_{s} \,,\qquad  N_{\rm sp}\simeq \left( \frac{ M_{\rm pl}}{M_s}\right)^{d-2}\,.
\end{equation}
Hence, the so defined species scale is equal to the string scale.
Note that here we have only exploited the presence of a string tower and the self-consistency of the relations \eqref{BekensteinH} and \eqref{entropystat}. This does not exclude that there might be a lower scale where the black hole  dynamically undergoes a phase transition. This was discussed in the context of transitions from towers of states to (minimal) black holes in \cite{Basile:2023blg} and was also employed for a conjecture on the characteristic energy scales appearing in an EFT of QG more recently in \cite{Bedroya:2024uva}.
Let us mention that one could try to use the relations \eqref{speciesscale} and \eqref{speciesnumber} to determine the species scale \cite{Marchesano:2022axe,Castellano:2022bvr,Blumenhagen:2023yws}, which leads to the result $\tilde\Lambda\simeq M_{s} \log(M_{\rm pl}/M_s)$.

The multiplicative $\log$-factor seems to be unphysical, as $\tilde\Lambda\simeq M_s$  is consistent with the known string corrections to the Einstein-Hilbert action, which include higher derivative terms generically suppressed by the string scale.
For the already mentioned K\"ahler moduli of type IIA compactifications on Calabi-Yau manifolds, it was argued in \cite{vandeHeisteeg:2022btw}  that the one-loop topological free energy $\mathbb{F}_1(T,\ov T)$ provides a good measure for the number of light species so that the species scale was proposed to be
\begin{equation}
\tilde\Lambda\simeq \frac{M_{\rm pl}}{\sqrt{ \mathbb{F}_1}}\,,
\end{equation}
which receives additive and not multiplicative corrections \cite{Cribiori:2023sch}.
One can show that for the aforementioned emerging string limit,
$\mathbb{F}_1\simeq t_B$ so that $\tilde\Lambda\simeq  M_s$.
Hence, we summarize that in an emergent string limit the
species scale coincides with  the string scale (of the emergent string)
and that there are no towers of states with a parametrically lighter mass.

Given the above generalization of the string scale as the scale where quantum effects of gravity become important in the presence of light towers of states, it is now straightforward to also apply it to the type IIA decompactification limit discussed in the previous section.
In this case, it is much shorter to employ the definitions \eqref{speciesscale} and \eqref{speciesnumber} for the computation of the species scale. 
The corresponding black hole computation was presented in \cite{Blumenhagen:2023yws} and gives the same result. 
Recall that the lightest states were BPS bound states of $D0$-branes leading to a tower with masses
\begin{equation}
M^{n}_{D0} \simeq \frac{M_s}{g_s} n\simeq \frac{M_{\rm pl}^{(d)}}{\lambda^{\frac{2(d-1)}{3(d-2)}}} n\,.
\end{equation}
The number of light species is given by the maximal KK mode, i.e~$N_{\rm sp}=n_{\rm max}\simeq\lambda^{\frac{2(d-1)}{3(d-2)}}\,\tilde\Lambda/M_{\rm pl}^{(d)}$ so that we can solve for
\begin{equation}
\tilde\Lambda\simeq \frac{M_{\rm pl}^{(d)}}{\lambda^{\frac{2}{3(d-2)}}}\simeq
M_{\rm pl}^{(d+1)}\simeq M_*\,.
\end{equation}                
As expected for  decompactification limits, the species scale is given by the Planck scale of one dimension higher, which here scales in the same way as  the 11D Planck scale. 
In the infinite distance limit this goes to zero, signalling that the $d$-dimensional theory breaks down and one has to describe the theory in $(d+1)$ dimensions.
However, we are not yet done. Since $\tilde\Lambda$ is parametrically larger than the mass scale of the $D0$-brane tower, there could still be other towers of states with a mass scale smaller than $\tilde\Lambda$.
These would further lower the species scale. 
However, the next lightest states are the aforementioned bound states of wrapped $D2$- and $NS5$-branes as well as KK modes along internal directions, whose mass scales precisely as $\tilde\Lambda$.  
Therefore, they do not further lower the species scale which is indeed given by the higher dimensional Planck scale.\footnote{An algorithm to calculate the species scale in the presence of multiple towers can be found in \cite{Castellano:2021mmx}.}

In \cite{Blumenhagen:2023xmk} a similar analysis was also done
  for the coscaled strong coupling limit of the 10D  type IIB superstring. As expected, in this
  case the lightest towers are the string towers of the $D1$-branes, so that the
  species scale is nothing else than the $D1$-string mass scale
  $\tilde\Lambda\simeq T_{D1}^{1/2}\simeq M_{\rm pl}/g_s^{1/4}$ with all
  other mass scales being parametrically larger than $\tilde\Lambda$.
  The analogous  result holds for the coscaled  type IIB limit in lower dimensions
  with the resulting species scale $\tilde\Lambda\simeq T_{D1}^{1/2}\simeq M_{\rm pl}^{(d)}/\lambda^{\frac{2}{d-2}}$.

\subsection{The Emergence Proposal}

We have seen that a genuine feature of QG is the existence of infinite distance limits which come in two different types, namely emergent string and decompactification limits. The QG cut-off is the species scale, which is the string scale for emergent string limits and the higher dimensional Planck scale for decompactification limits.
In these limits towers of states become asymptotically massless and one has a naturally small parameter in which one can hope to formulate perturbation theory.
The prime example is (fundamental) string theory itself, where this parameter is just the string coupling, $g_s=\exp(\phi)$. 

In this context, in \cite{Heidenreich:2017sim,Grimm:2018ohb,Heidenreich:2018kpg} an interesting observation was made, namely that the metric on moduli space can be recovered by integrating out the tower of asymptotically massless states at one-loop. 
As a simple toy model, consider a light modulus $\phi$ and a tower of massive KK states $h_n$ with mass $M_n=n\, \Delta m(\phi)$  governed by a $d$-dimensional effective action
\begin{equation}
\label{towereffact}
S=M_{\rm pl}^{d-2} \int d^d x \left( \frac{1}{2} G_{\phi\phi}\, \partial_\mu\phi  \partial^\mu \phi+ \sum_n  \frac{1}{2}  \partial_\mu h_n \partial^\mu h_n +  \frac{1}{2} m^2_n(\phi)  h_n^2 \right)\,,
\end{equation}
where $G_{\phi\phi}$ denotes the metric on field space.
The moduli-dependent mass terms for $h_n$ leads to  three-point
couplings $y=[m_n (\phi) \partial m_n(\phi)]\,  h_n^2\, \phi$, inducing
a one-loop correction to the
kinetic term for $\phi$ with the KK modes $h_n$  running in the loop. 
Integrating out these modes up to the UV cut-off, which is taken to be
the species scale, leads to the leading order  one-loop correction (see e.g.~\cite{Castellano:2022bvr} for more details)
\begin{equation}
\label{metriconeloop}
G_{\phi\phi}^{\rm 1-loop} \simeq  \frac{\tilde\Lambda^{d-1} }{ M_{\rm pl}^{d-2}} \frac{  \left( \partial_\phi \Delta m(\phi) \right)^2 }{\left( \Delta  m(\phi) \right)^{3}}+\ldots\,.
\end{equation}
For a KK tower with $\Delta m=M_{\rm pl}/r$, the species scale is the $(d+1)$-dimensional Planck scale, i.e.~$\tilde\Lambda^{d-1} \simeq M_{\rm pl}^{d-1}/r$.
Then we find $ G_{rr}^{\rm 1-loop}\simeq1/r^2$, which has the same functional dependence on the modulus $r$ as the tree level metric, $G_{rr}^0$, resulting  from the dimensional reduction of the Einstein-Hilbert action.
Even though this is just a simple toy model, the above was considered quite a remarkable correlation leading to the formulation of the so-called Emergence Proposal:

\begin{quotation}
  \noindent
  {\it  Emergence Proposal (Strong):
The dynamics (kinetic terms) for all fields are emergent in the infrared by integrating out towers of states down from an ultraviolet scale $\Lambda_s$, which is below the Planck scale.}
\end{quotation}

\noindent
There have been slightly different formulations and also a weak version \cite{Castellano:2022bvr}, but
for the purpose of this presentation let us stick to this version
formulated in the review \cite{Palti:2019pca}.

As it stands, this   proposal is very generic and it is not clear what its realm of validity could be. Of course, it is certainly meant in the context of QG and the swampland program, where one usually identifies the UV cut-off scale $\Lambda_s$ with the species scale, i.e. $\Lambda_s\simeq \tilde\Lambda$.
Moreover, the towers of states to be integrated out are those described in the previous sections and which become light in infinite distance limits of the moduli space.

In the previous toy example, a hard UV cut-off for the one-loop integral was introduced, i.e~ both internal momenta and the mass of the states running in the loop were cut-off at the species scale. 
However, when extending this proposal beyond the pure EFT setting to theories of QG, the example of the fundamental string tells us that one should not cut-off the loop-integrals at a finite energy scale and keep only the string modes with masses below that scale. In fact, such string loop amplitudes have nice UV properties precisely by including all infinitely many states from the tower, as only then we have modular invariance and we can restrict the integration over the fundamental domain of $SL(2,\mathbb Z)/\mathbb Z_2$. 
Hence, despite the treatment of the simple toy example, in the Emergence Proposal it is implicitly meant that one really integrates out the full infinite tower of states. This is also compatible with  the calculations of \cite{Palti:2019pca} (see footnote $^{46}$ therein).\footnote{We thank E. Palti for confirming this point.}

Then the question of which towers one has to integrate out arises: are they only the lightest one or even all conceivable towers?
The latter option can be excluded, as in weakly coupled string theory, one only integrates out those towers with a mass scale $M_s$. 
As mentioned at the beginning of this section, $p$-brane towers are considered as classical non-perturbative objects and are not running in the loop. 
In the next section, we will provide evidence that for decompactification limits  one also has to integrate out more than just the lightest tower. 
Hence, we think that the answer is likely in the middle and one has to integrate out all full infinite towers of states with mass scale not larger than the species scale. 
In analogy to string theory, these will be considered as the perturbative states in the effective description of QG, while all the heavier towers of states will be classical and non-perturbative. 
This means that we identify the scale $\Lambda$ from \eqref{pertandclass} with the species scale, i.e~$\Lambda= \tilde\Lambda$.
In addition, while the Emergence Proposal explicitly mentions kinetic terms, one could conceive that in a fully emerging effective theory also all higher derivative terms are generated by quantum effects.

The emergent string conjecture tells us that there are only two different kinds of infinite distance limits: the emerging string limit and the decompactification limit. Can the Emergence Proposal be true for an emerging string limit? 
We think that the answer is negative for the following reasons. 
First, we notice that even the naive computation around \eqref{towereffact}-\eqref{metriconeloop} for a string tower does not give correct leading order results, as the previously mentioned multiplicative $\log$-terms in the species scale are transferred to the one-loop corrections \cite{Blumenhagen:2023yws}. 
Second and at a more fundamental level, we know how to quantize the weakly coupled fundamental string and in fact, similarly to QFT, one obtains a loop expansion in terms of higher genus Riemannian surfaces. 
Here, e.g.~the one-loop Schwinger integral over the tower of string states gives (tautologically) only the one-loop correction to certain terms in the low energy effective action. Hence, none of the tree-level terms in the $g_s$ expansion are generated from quantum effects. 
This will also be true for other emerging string limits, like the one discussed in  the type IIA K\"ahler moduli space for a $K3$-fibered Calabi-Yau, as these are conjectured to be dual to a fundamental string.
We refer to \cite{Blumenhagen:2023tev} for more details.
The exclusion of weakly coupled string limits resonates with the pragmatic concept of emergence described at the beginning of this article.

With the emergent string limit excluded, it is only the decompactification limit that remains, of which the M-theory limit is the typical example and perhaps the only non-trivial one. In this respect, we note that in
  standard   (non-coscaled) decompactification limits $R\to\infty$ with $g_s\ll 1$ and  the volume ${\cal V}$ (in string units)  of the orthogonal $9-d$ dimensional compact space held fixed, the perturbative  quantum gravity theory still contains light strings. Certainly, the lightest tower of states
  is the KK tower with $m_{\rm KK}\sim 1/R$, whose induced species scale is
  the finite $(d+1)$-dimensional Planck scale.
  However, this is related to the string scale via
  \eq{
 \tilde\Lambda\simeq M_s\left(\frac{\mathcal{V}}{g_s^2}\right)^{\frac{1}{d-1}}\,,
    }
 which  for   $g_s\ll 1$ and ${\cal V}>1$  is larger than the string scale $M_s$.
Therefore, as expected, such a limit is just a higher dimensional perturbative string theory. Even though the lightest states are given by KK towers, the QG theory is described by quantized strings and
as for the aforementioned emergent string limit,
  the Emergence Proposal is not realized.\footnote{In fact, there have been examples of the emergence proposal not being straightforwardly realized in such decompactification limits, like the partial emergence of certain quartic gauge couplings analyzed in \cite{Lee:2021usk}.}

As we have seen, the M-theoretic decompactification limit is of a different type as string towers are heavier
than the species scale.
The QG theory of M-theory is arguably one of the deepest mysteries and only partial results are available at present, like a formulation in terms of $D0$-branes, the BFSS matrix model \cite{Banks:1996vh} (see
\cite{Bilal:1997fy,Bigatti:1997jy,Taylor:2001vb} for reviews). 
While later on we will present a more detailed discussion, we can already state that in the BFSS matrix model the interaction between gravitons was indeed found to be absent classically and  only generated via quantum (loop) effects. 
This can indicate that there is a good chance for the M-theory limit to be the natural home of the Emergence Proposal.
In this spirit, from our discussion we extrapolate a lesson in the form of an M-theoretic refinement of the Emergence Proposal:

\begin{quotation}
\noindent
{\it  Emergence Proposal (M-theory):
In the  infinite distance M-theory limit $M_* R_{11}\gg 1$ with the Planck scale kept fixed,  a perturbative QG theory arises whose low energy effective description emerges via quantum effects by integrating out the full infinite towers of states with a mass scale parametrically not larger than the 11D Planck scale. 
These are transverse $M2$-, $M5$-branes carrying momentum along the eleventh direction ($D0$-branes) and along any potentially present compact direction.}
\end{quotation}

\noindent
Note that in this limit the longitudinally wrapped $M2$-brane, i.e.~the type
IIA fundamental string, and  the longitudinally wrapped $M5$-brane, i.e.~the type
IIA $D4$-brane, have masses
\begin{equation}
M_{F1} = \frac{ M_*}{g_M^{1/2}}\,,\qquad\qquad  M_{D4} = \frac{ M_*}{g_M^{1/5}}\,, 
\end{equation}
with the formal coupling constant $g_M=1/(M_* R_{11})\ll 1$.
Hence, they are not among the light modes and are considered
as non-perturbative classical objects.
We note that the BFSS matrix model is indeed containing
these two longitudinal branes  as bound states of $D0$-branes,
which one might speculate to be the analogue of the description
of non-perturbative $D$-branes as coherent (boundary) states
of weakly coupled closed string modes. 
However, the transverse $M5$-brane was lacking in the original BFSS matrix model which in view of the Emergence Proposal might indicate that
it is not yet the complete description of quantum M-theory.\footnote{Transverse $M5$-branes appear in the BMN version of the BFSS matrix model \cite{Maldacena:2002rb} and, more recently, they have been investigated in cohomotopy and in connection to ``Hypothesis H'' \cite{Sati:2019nli,Corfield:2021wwd}.}
Before delving too deeply into such speculations, we provide more evidence for our M-theoretic Emergence Proposal.

\section{Evidence for the M-theoretic Emergence Proposal}
\label{sec:emMth}

The central challenge is to provide evidence despite the obvious shortcoming that we do not understand the full quantization of M-theory, yet. In addition,  as M(matrix) theory teaches us, namely that the leading order supergravity action at the second  derivative level is emerging via loops, also space-time itself should be somehow emergent.

The loophole bypassing these difficulties is that there are certain couplings in the effective action that are protected by supersymmetry and do only receive contributions from 1/2 BPS states. 
These states are under good control and, up to a certain extend, can already be reliably described by their weak string coupling counterparts,
i.e.~in the weakly coupled type IIA theory.
Hence, this sector of M-theory is special and admits the usual geometric interpretation we are used to from string theory. 
We will see that indeed the string one-loop evaluation for 1/2 BPS states can be extended to M-theory, providing very reasonable results.

This is reminiscent of the working extension from  Double Field Theory (DFT) to Exceptional Field Theory (ExFT) (see \cite{Aldazabal:2013sca,Hohm:2013bwa,Berman:2020tqn} for reviews). 
The section conditions are in fact the 1/2 BPS conditions for the corresponding $M2$-, $M5$-branes and KK modes, written as differential operators on the extended space made from usual and (brane) winding coordinates \cite{Bossard:2015foa}.
In  DFT and ExFT one also truncates the complete spectrum to just KK and brane-wrapping modes leaving out the string, respectively M-theory, excitations.

In theories with 32 supercharges, the higher derivative $R^4$-term is 1/2 BPS saturated. Longer supermultiplets, preserving less supersymmetry, do not contribute to it (see e.g.~\cite{deWit:1999ir}).
This term has received attention lately in the context of species scale calculations \cite{vandeHeisteeg:2023dlw,Castellano:2023aum} and of the emergence of species scale black hole horizons \cite{Calderon-Infante:2023uhz}.
In the former, the  coefficient of the $R^4$-term has been shown to give the expected cut-off for emergent string and decompactification limits, i.e.~the string scale and the higher dimensional Planck mass respectively.
In theories with 16 supercharges, like type IIA on $K3$ or the dual heterotic string on $T^4$, the $F^4$-coupling is 1/2 BPS saturated.
In theories with 8 supercharges, like $N=2$ supergravity in 4D,   the topological string couplings ${\cal F}_g$ at arbitrary genus $g$ are 1/2 BPS saturated. 
Note that ${\cal F}_0$  contains information about the second order supergravity action, namely about the gauge couplings and kinetic terms for the K\"ahler moduli.
In this sense, such a coupling is special and might be sensitive to issues related to the emergence of space-time itself.

\subsection{Emergence of \texorpdfstring{$R^4$}{TEXT}-terms}

Let us start with a theory featuring  maximal supersymmetry, i.e.~the 10D type IIA superstring compactified to $d$ dimensions on a $k$-dimensional torus.
The higher derivative $R^4$-term arises in the low-energy effective action as
\begin{equation}
\label{R4termd}
S_{R^4}\simeq M^{d-8}_s\,  V_k\int d^d x  \sqrt{-g}\, a_d\, t_8 t_8\, R^4\,,
\end{equation}
where  $g_{\mu\nu}$ denotes  the string frame metric and $V_k$ the volume of the internal torus in string units.
For simplicity, we restrict ourselves to rectangular tori and set $B_2=0$.
In the emergent string limit, i.e.~for small $g_s$, the coefficient $a_d$ receives at most  one-loop perturbative corrections and space-time instanton corrections. 
The general form can be schematically written as
\begin{equation}
a_d = \frac{c_0}{g_s^2} +\underbrace{\left(c_1 +{\cal O}\left(e^{-S_{\rm ws}}\right) \right)}_{\rm one-loop} + {\cal O}\left(e^{-S_{\rm st}}\right) \,,
\end{equation}
where $S_{\rm ws}$ denotes the action of world-sheet instantons and
$S_{\rm st}$ that of space-time instantons.
The tree-level and one-loop coefficients are known to be
$c_0=2\zeta(3)$ and $c_1=2\pi^2/3$ where $\zeta(s)=\sum_{n=1}^{\infty}
n^{-s}$ is the Riemann zeta function.
Our task is to compute $a_d$ in the  M-theory limit, where in particular $g_s \gg 1$. Since the coupling is 1/2 BPS, in both limits one should get the same result and the Emergence Proposal claims that, in the M-theory limit,  $a_d$ should stem entirely from  quantum effects without any classical contribution.

Before discussing the M-theory limit, it is worthwhile to recall a few aspects of the one-loop computation for $a_d$ in the weakly coupled type IIA string. This will help us to sharpen our technical tools and to understand certain analogies between the string and the M-theory computation.

\subsubsection{Emergent string limit}

An essential step of computing the one-loop diagram is a proper regularization method for the real Schwinger integrals that are naively UV divergent, such as
\begin{equation}
\log(p^2 +m^2)\sim \int_0^\infty \frac{dt}{t} e^{-\pi t (p^2 + m^2)}\,.
\end{equation}    
In string theory it is well established that the real Schwinger parameter $t$ is complexified to $\tau=\theta+it$ by implementing the string level-matching condition,
\begin{equation}
  L_0-\ov{L}_0=m_i\,  n^i +N-\ov N=0\,,
\end{equation}  
via a Lagrange multiplier $\theta$ and then using modular invariance to restrict the complex integration to the fundamental domain $\mathcal{F}$ of $SL(2,\mathbb Z)$, thus avoiding the UV singularity. 
Notice that for vanishing string excitations, $N=\ov N=0$, the level matching condition becomes the 1/2 BPS condition $m_i n^i=0$ for KK momenta $m_i$ and
string winding modes $n^i$. 
Proceeding in this manner, the one-loop contribution to the coefficient of the $R^4$-term in $d$ dimensions can be expressed as
\begin{equation}
\label{adGGVstring}
a_{d,{\rm string}}^{(1)}  \simeq \frac{2\pi}{{V}_k} \sum_{m_i, n^i \in \mathbb{Z}}  \int_{\cal F} \frac{d^2\tau}{\tau_2^{\frac{d-6}{2}}}   \, e^{-\pi \tau_2 M^2 -2\pi i \tau_1 m_i n^i}\,,
\end{equation}   
with
\begin{equation}
M^2= m_i G^{ij} m_j + n^i G_{ij} n^j\, ,
\end{equation}   
and $G_{ij}$ the metric on the torus.      
In this expression the Gaussian integral over the continuous momenta along the $d$ non-compact directions has already been carried out.
Going one step back, this is the stringy regularization of the initially UV divergent expression
\begin{equation}
\label{adGGVdiver}
a_{d}^{(1)} \simeq \frac{2\pi}{V_k} \sum^{\_\_\_}_{m_i, n^i \in \mathbb{Z}} \int_{0}^{\infty} \frac{dt}{t^\frac{d-6}{2}}   \, \delta({\rm BPS}) \, e^{-\pi t M^2 }\,,
\end{equation}
which is sort of not taking into account the extended nature of the
string, and hence is affected by UV divergence close to $t=0$.
In \eqref{adGGVdiver}, the symbol  $\overline\sum$ is denoting the sum with the term with all $m_i$ and $n^i$ vanishing excluded, so that the expression is related to the definition of constrained Eisenstein series in \cite{Obers:1999um},
while $\delta({\rm BPS})$ arises from carrying out the unfolded integral over $\theta$ using $\delta(x) = \int_{-\infty}^{\infty} d\theta e^{-2\pi i x \theta}$.

Let us consider the simplest case, $d=10$. Evaluating the string
integral \eqref{adGGVstring} gives the finite result
\begin{equation}
a_{10,{\rm string}}^{(1)}  \simeq 2\pi \int_{\mathcal{F}} \frac{d^2\tau}{\tau_2^2}=\frac{2\pi^2}{3}\,.
\end{equation}
On the other hand, introducing a UV cut-off $\epsilon >0$ in the divergent integral \eqref{adGGVdiver},
one can write
\begin{equation}
a_{10}^{(1)}  \simeq 2\pi \int_\epsilon^\infty \frac{dt}{t^2} = \frac{2\pi}{\epsilon} \,.
\end{equation}
One could just minimally subtract this term, but then in 10D one would arrive at $a_{10}^{(1)}=0$. 
Baring this limitation of \eqref{adGGVdiver} in mind, let us have a look at the $d=9$ case, where the wrapped strings are particle-like. 
Here we have KK and winding modes with mass $M^2=m^2/\rho^2+n^2 \rho^2$, with $\rho$ the radius of the circle in string units.
The evaluation of \eqref{adGGVstring} yields
\begin{equation}
\label{9dstringr4}
a_{9,{\rm string}}^{(1)} \simeq \frac{2\pi^2}{3} \left( 1 + \frac{1}{\rho^2} \right)\,.
\end{equation}
Now let us try again to regularize the divergent integral \eqref{adGGVdiver}.
From the BPS condition $m \cdot n = 0$ we see that two different sectors contribute, namely one with only winding, $n\ne 0$, and the other with only KK momentum, $m\ne 0$.  
We proceed as in the previous 10D case and introduce a UV regulator $\epsilon>0$ to get 
\begin{equation}
\int_{\epsilon}^{\infty}\frac{d t}{t^{3/2}}e^{-\pi t A}=\frac{2}{\sqrt\epsilon}-2\pi\sqrt{A}+\mathcal{O}(\sqrt{\epsilon})\,,
\end{equation}
where we expanded around $\epsilon\simeq 0$. 
We regularize this expression via minimal subtraction of the divergent term, $2/\sqrt{\epsilon}$, and by sending $\epsilon\to 0$ afterwards. 
Thus, the winding sector contribution becomes
\begin{equation}
\label{a9m=0}
a_{9,m=0}^{(1)} \simeq \frac{2\pi}{\rho} \sum_{n \neq 0} \int_{0}^{\infty} \frac{dt}{t^{3/2}}\,  e^{-\pi t \rho^2 n^2} = -4\pi^2 \sum_{n\ne 0} |n|  =   \frac{2\pi^2}{3}\,,
\end{equation}
where we regularized the sum over $n$ via the zeta-function, $\zeta(-1)=-1/12$. Following the same procedure for the KK contribution one obtains
\begin{equation}
\label{a9n=0}
a_{9,n=0}^{(1)} \simeq \frac{2\pi}{\rho} \sum_{m \neq 0} \int_{0}^{\infty} \frac{dt}{t^{3/2}} \,e^{-\pi t\frac{ m^2}{\rho^2}} = \frac{2\pi^2}{3} \frac{1}{\rho^2}\,,
\end{equation}
so that by combining \eqref{a9m=0} and \eqref{a9n=0} the full string one-loop result \eqref{9dstringr4} is recovered.
In \cite{Blumenhagen:2024ydy} this regularization method was applied also for lower non-compact spacetime dimensions and was shown to give consistent results.
For later purposes, we provide here the result in 8D
\begin{equation}
a^{(1)}_{8}\simeq  - \frac{2\pi}{T}  \log\left(T\,  \left|\eta(i T) \right|^4 \right)  -\frac{2\pi}{T}  \log\left(U\,  \left|\eta(i U)   \right|^4 \right)\,,
\end{equation}
with $T=\rho_1\rho_2$ and $U=\rho_2/\rho_1$.  
Notice that the first term  shows the presence of world-sheet instantons.

To summarize, via minimal subtraction of the UV divergence and zeta-function regularization of the infinite sums, we have found an alternative way to regularize the divergent real Schwinger integral giving the same result as the known regularization performed in string theory. 
Only in 10D the method does not apply directly, and we believe this to be related to the fact that the generic winding string is not particle-like. 
Nevertheless, one can recover the correct 10D result from the 9D one via decompactification, $\rho\to\infty$, so that in principle the full information about the one-loop correction to the $R^4$-term can be computed via the regularization of the divergent expression \eqref{adGGVdiver} just mentioned. In the same  vein, one could calculate the one-loop contributions
in other emergent string limits, like e.g. the coscaled  strong coupling limit of type IIB.
Here one would expect that a Schwinger integral gives  the one-loop corrections in  $g_E=1/g_s\ll1$  to the $R^4$-coupling.

\subsubsection{Decompactification limit}

The question is if and how one can compute the $R^4$-coefficient $a_d$ also in the M-theoretic decompactification limit, $r_{11}=M_* R_{11}\gg 1$.  
As we will argue, the method for the evaluation of the real Schwinger integral introduced previously will be very useful in this regard.

Let us recall that, in this limit of strong type IIA coupling, swampland arguments suggest that we need to integrate out the towers of states with mass scale below the species scale, which is the 11D Planck scale.
These are transverse $M2$- and $M5$-branes carrying KK momentum along all compact directions, including the very large eleventh direction. 
For $d\le 3$ ($k\ge 7$), also the transverse KK-monopole should be taken into account. For simplicity, here we do not consider it and restrict ourselves to $d\ge  4$.
Note that the light modes do not include the type IIA fundamental string, i.e.~the longitudinal $M2$-brane, so that it is a priori  non-trivial that the former perturbative (in $g_s$) one-loop correction $a_d^{(1)}$ can be recovered.

In the pioneering work of Green-Gutperle-Vanhove (GGV) \cite{Green:1997as}, such a computation was performed for the first time. 
The coefficients $a_d$ in 10D and 9D were given by a natural generalization of the weakly coupled one-loop string formula \eqref{adGGVdiver}, where one was summing over the KK spectrum along the eleventh direction, i.e.~bound states of $D0$-branes.
This was generalized in \cite{Obers:1999um} (see also the closely related
work \cite{Green:1997di,Kiritsis:1997em,Russo:1997mk,Pioline:1997pu,Obers:1998fb}) to include the (full) 1/2 BPS particle-like states of M-theory in $d$ dimensions. 
The final expression for the one-loop Schwinger integral in perturbative M-theory can be compactly written as
\begin{equation}
\label{fullr4termdef}
a_{d,{\rm M}}^{(1)}\simeq \frac{2\pi}{r_{11}{\cal V}_{k}}   \sum^{\_\_\_}_{N^I,m\in\mathbb Z}  \int_0^\infty  \frac{dt}{t^{\frac{d-6}{2}}}\;  \delta({\rm BPS}) \;e^{-\pi t\, N^I  {\cal M}_{IJ}  N^J -\pi t \,\frac{m^2}{r_{11}^2}}\,,
\end{equation}
where all volumes and masses  are measured in M-theory units. 
Apparently, one now also integrates out the KK momentum $m$ along the eleventh direction, which has been isolated from the rest. 
We have  collectively denoted the transverse KK momenta $m_i$, $i\in\{1,\ldots, k\}$, and the various $M$-brane wrapping numbers as
\begin{equation}
N^I=\left(m_i, n^{ij}, n^{ijklm}\right)\,.
\end{equation}      
The mass matrix for 1/2 BPS states is given by
\begin{equation}
{\cal M}={\rm diag}\left(\frac{1}{r_i^2}, t_{ij}^2, t_{ijklm}^2\right)\,,
\end{equation}
where, as throughout the paper, we made the simplifying assumption of a rectangular torus with vanishing axionic fields. 
Turning on the latter induces off-diagonal terms in the mass matrix.
The quantity $t_{ij}=r_i r_j$ (and similarly for $t_{ijklm}$) denotes the volume wrapped by the corresponding transverse $M2$-brane (and $M5$-brane). 
Finally, the 1/2 BPS conditions involving KK modes and the transverse $M2$- and $M5$- wrapping numbers read \cite{Obers:1998fb}
\begin{align}
  &n^{ij} m_j=0\,,\hspace{3.25cm} \#=k\,,\\
  &n^{[ij}\, n^{kl]} + m_p \,n^{pijkl}=0\,, \hspace{1cm} \#={\binom{k}{4}}\,,\\
  & n^{i[j} \,n^{klmnp]}=0\,, \hspace{2.28cm}  \#=k\binom{k}{6}\,,
\end{align}
where we also indicated their number.
The first condition says that the momentum has to be orthogonal to the world-volume of the $M2$-brane. For vanishing $M5$-brane wrapping number, the second condition means that the matrix $n^{ij}$ has rank two.

Before presenting the evaluation of this compact expression for  $a_{d,{\rm M}}^{(1)}$ in a few examples, let us note that there is a group-theoretic structure behind the number of particle-like states and their 1/2 BPS conditions, which is closely related to a similar structure in exceptional field theory. 
Following \cite{Obers:1999um}, let us collect all light transverse particle states that we integrate out in the Schwinger integral.
These states form bound states with the unrestricted $D0$-branes, i.e.~the KK modes along the eleventh direction, and fit nicely into representations of $E_{k(k)}(\mathbb Z)$, which we denote as $E_{k(k)}$ to simplify our notation. 
We define as usual $E_{2 (2)}=SL(2)$, $E_{3 (3)}=SL(3)\times
SL(2)$, $E_{4 (4)}=SL(5)$ and $E_{5 (5)}=SO(5,5)$. 
In table \ref{table_particlesinE}, we list all particle states and how they fit
into representations $\Lambda_{E_k}$ of  $E_{k(k)}$ as well
as the representations $\lambda_{E_k}$ of the 1/2 BPS conditions (the latter coinciding with the representation of the string multiplet \cite{Pioline:2010kb}).

\vspace{0.3cm}
\begin{table}[ht] 
\renewcommand{\arraystretch}{1.5} 
\begin{center} 
\begin{tabular}{|c|c|c|c|c|c|} 
\hline
d & k & Particles $SL(k)$ reps. &  $E_{k(k)}(\mathbb Z)$ &  $\Lambda_{E_k}$  & 1/2 BPS: $\lambda  _{E_k}$      \\
\hline \hline
9 & 1 & $[1]_p$  &  $1$ & 1 & 0  \\
8 & 2 & $[2]_p + [1]_{M2}$  &  $SL(2)$ & 3 & 2  \\
7 & 3 & $[3]_p + [3]_{M2}$  &  $SL(3)\times SL(2)$ & (3,2) & (3,1)  \\
6 & 4 & $[4]_p + [6]_{M2}$  &  $SL(5)$ & 10 & 5  \\
5 & 5 & $[5]_p + [10]_{M2}+[1]_{M5}$  & $SO(5,5)$  &  16 & 10  \\
4 & 6 & $[6]_p + [15]_{M2}+[6]_{M5}$  &   $E_6$ & 27 & 27  \\
\hline
\end{tabular}
\caption{Particle states, 1/2 BPS conditions and their $E_k$ representations for $k\le 6$. $[k]_p$ denotes KK momenta along the $k$ transverse directions.}
\label{table_particlesinE}
\end{center} 
\end{table}

\begin{figure}[ht]
\centering
\vspace{-1.7cm}
\includegraphics[width=\textwidth]{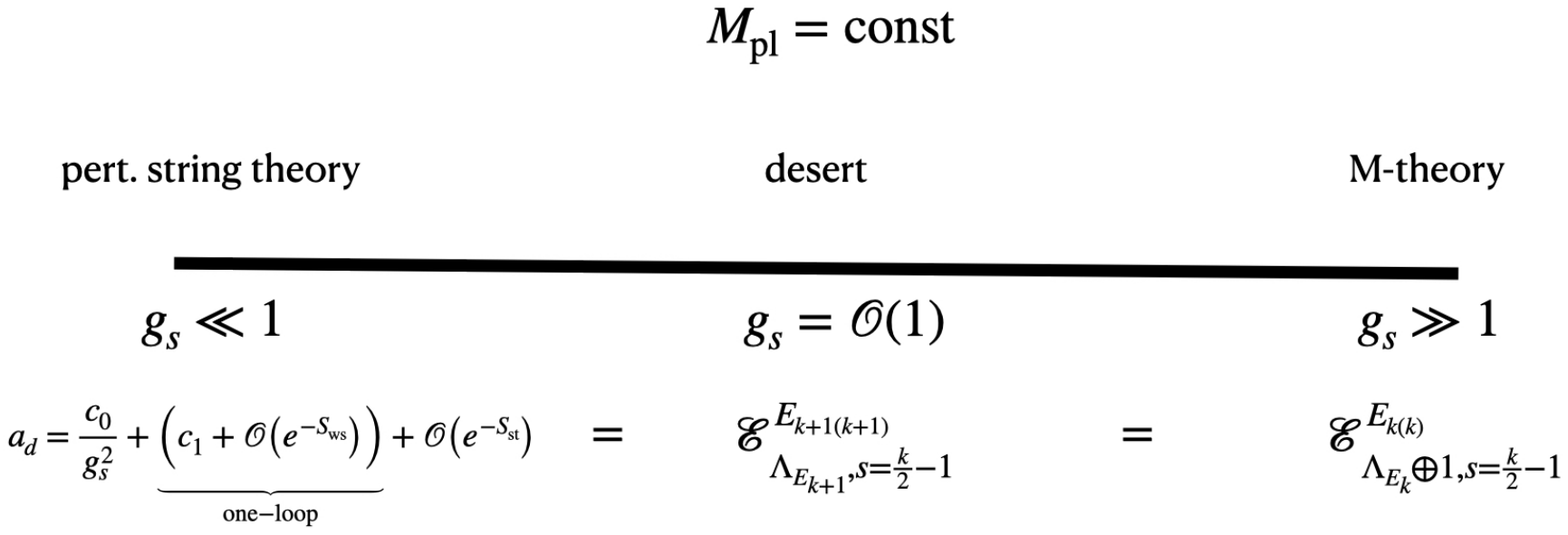}
\vspace{-3.3cm}
\caption{Schematic view of the coscaled  dilaton moduli space.}
\label{fig_moduli}
\end{figure}

Due to this structure, the Schwinger integral \eqref{fullr4termdef} may be viewed as a constrained Eisenstein  series
\begin{equation}
  a_{d,{\rm M}}^{(1)}= {\cal E}^{E_{k(k)}}_{\Lambda_{E_k}\oplus 1, s={\frac k2}-1}\,.
\end{equation}
Notice that there is a shift by one dimension relatively to the full U-duality group  $E_{k+1(k+1)}$ of toroidal compactifications of M-theory on $T^{k+1}$ down to $d$ dimensions. The latter was the guiding principle in \cite{Obers:1999um}, where all 1/2 BPS states were considered, including also the longitudinal $M$-branes, which are instead excluded in our counting. 
In particular, in \cite{Obers:1999um} the resulting total Schwinger integral is identified with a constrained Eisenstein series of the type
\begin{equation}
\label{constrEins}
a_{d,{\rm bulk}}^{(1)}={\cal E}^{E_{k+1(k+1)}}_{\Lambda_{E_{k+1}}, s={\frac k2}-1}\,,
\end{equation}
with  $d+k=10$ for $k>2$.
In other words, since we are consistently working in the large $r_{11}\gg 1$ region the full U-duality group $E_{k+1(k+1)}$ is broken to a subgroup
$E_{k(k)}\times 1$, which distinguishes the eleventh direction. As depicted in figure \ref{fig_moduli},
physically the difference between our approach and \cite{Obers:1999um} is
that we are consistently working in the perturbative
decompactification limit
where we only integrate out the perturbative states, i.e.~transverse
$M$-branes, whereas in  \cite{Obers:1999um}  the Schwinger integral involves
all states, meaning that they work in the interior (or the desert) of the $r_{11}$ moduli space. 
Due to supersymmetric protection, in the end both results for the $R^4$-term should be equal, as we have shown explicitly for $d\ge 8$ in \cite{Blumenhagen:2024ydy}, while a more general
proof is available for $d\ge 4$ in the previous work \cite{Bossard:2015foa,Bossard:2016hgy}. 

We notice that in the bulk region, $g_s =\mathcal{O}(1)$, of the type IIB
  superstring, the complete  $R^4$-coupling can also be derived from the M-theory
  expression  \eqref{constrEins}  for $d\le 9$ by applying the map \eqref{map2bm}
 between M-theory and type IIB quantities. As usual, the 10D decompactification limit of type IIB
 is given by the $t=r_1\, r_{11}\to 0$ limit in M-theory.
 Hence, in the bulk one always deals with M-theory and the $R^4$-couplings are emerging by integrating out all 1/2-BPS states, including transverse and longitudinal ones. This is certainly the least
 understood regime of QG, where the species scale is of the same order as
 the  Planck scale. The lesson one might draw from this is that this genuine QG theory is expected to
 feature emergence of space-time and all its interactions. From this perspective,
 the all familiar emergent string limits are special in that they admit a
 weakly coupled perturbation theory, still fairly analogous to the description of perturbative  QFTs.

\subsubsection*{Emergence of $R^4$-term in 10D}

Let us evaluate the expression \eqref{fullr4termdef} in 10D, i.e. for $d=10$ and $k=0$.
As already mentioned,  the coefficient of the $R^4$-term contains only a string tree-level and a one-loop term,  
\begin{equation}
\label{r410dknown}
a_{10}\simeq \frac{2 \zeta(3)}{g_s^2} +\frac{2\pi^2}{3}\,.
\end{equation}
The M-theoretic Schwinger integral involves only a sum over KK modes and hence simplifies considerably to
\begin{equation}
\label{r410dschwinger}
a^{(1)}_{10,{\rm M}}\simeq \frac{2\pi}{ r_{11}}  \sum_{m\ne 0} \int_0^\infty  \frac{dt}{t^{2}}  \;  e^{-\pi t \frac{m^2}{r_{11}^2}}\,.
\end{equation}
As in the string case, this real integral is divergent for $t\to 0$, so that we proceed with its regularization along the same line as for the divergent real string-theoretic Schwinger integrals \eqref{adGGVdiver}.
First, we introduce a regulator $\epsilon>0$ and  perform a minimal subtraction of the divergence to arrive at
\begin{equation}
a^{(1)}_{10,{\rm M}}\simeq \frac{2\pi^2}{r_{11}^3} \sum_{m\ne 0} m^2 \log\left(\frac{m^2}{\mu^2}\right)\,,
\end{equation}
with a constant $\mu^2$.
Next, the sum over $m$ is carried out employing zeta-function regularization.
Using $\zeta(-2)=0$, the $\mu$-dependent term drops out and what remains can be expressed as
\begin{equation}
\label{result10D}
a^{(1)}_{10,{\rm M}}\simeq \frac{2 \zeta(3)}{r_{11}^3} =  \frac{2\zeta(3)}{g_s^2}\,,
\end{equation}
where we employed
\begin{equation}
\sum_{m\ge 1} m^2 \log(m)=-\zeta'(-2)=\frac{\zeta(3)}{4\pi^2}\,.
\end{equation}
Thus, the Schwinger integral for the tower of $D0$-branes reproduces precisely the tree-level term in the expansion of  \eqref{r410dknown} at small $g_s\,$.
This result was already obtained in the original work of GGV \cite{Green:1997as}, but can now be interpreted as first evidence for the M-theoretic Emergence Proposal. 

However, the string one-loop term in \eqref{r410dknown} is missing in
our computation above.\footnote{In this respect, we recall that GGV added an (infinite) constant $C$ that was fixed in a subsequent step to the correct value $2\pi^2/3$ by invoking T-duality in 9D.}
As we will see, this is completely analogous to the missing $2\pi^2/3$ term in the 10D real Schwinger integral \eqref{adGGVdiver}. More specifically, the object that generates it is not particle-like in 10D. Let us mention that in the $g_s\gg 1$ limit, the one-loop term in \eqref{r410dknown} is leading over the tree-level term.

\subsubsection*{Emergence of $R^4$-term in 9D}

For $d=9$ and $k=1$, the Schwinger integral \eqref{fullr4termdef} with still only KK contributions is given by
\begin{equation}
\label{r49dschwinger}
a^{(1)}_{9,{\rm M}}\simeq \frac{2\pi}{r_{11} \,r_{1}}
\sum_{(m,n)\ne (0,0)} \int_0^\infty  \frac{dt}{t^{3/2}}  \;  e^{-\pi t\left(  \frac{m^2}{r_{11}^2}+\frac{n^2}{r_{1}^2}\right)}\, .
\end{equation}
In order to compare the result to weakly coupled type IIA, one has to apply a couple of steps to eventually obtain an expression that admits an interpretation in terms of string loop and instanton corrections.
For that purpose, one first splits the sum according to $\sum_{(m,n)\ne (0,0)} =\sum_{m=0,n\ne 0}+\sum_{m\ne 0,n\in \mathbb Z}\,$.
The first piece ($m=0$) can be treated straightforwardly to obtain
\begin{equation}
\label{r49dfirst}
 a^{(1)}_{9,{\rm M};(m=0)}\simeq \frac{2\pi^2}{3}\frac{1}{r_{1}^2 \,r_{11}}
 \,.
\end{equation}  
For the second piece ($m\neq 0$), one performs a Poisson resummation of the sum over $n\in\mathbb Z$ to obtain
\begin{equation}
a^{(1)}_{9,{\rm M};(m\neq 0)}\simeq \frac{2\pi}{r_{11}} \sum_{m\ne 0} \sum_{n\in\mathbb Z}  \int_0^\infty  \frac{dt}{t^{2}} \;e^{ -\pi t \frac{m^2}{ r_{11}^2} -\frac{\pi}{t} n^2 r_{1}^2}\,.
\end{equation}
The $n=0$ term gives  precisely the sum \eqref{r410dschwinger} from the previous section, that is equal to $2\zeta(3)/r_{11}^3$, while for the remaining sum over $n\ne 0$ one employs the relation
\begin{equation}
\label{besselrel}
\int_0^\infty \frac{dx}{x^{1-\nu}} \,e^{-{\frac{b}{x}}-cx}=2 \left|  {\frac{b}{c}}\right|^{\frac{\nu}{2}} K_\nu\left(2\sqrt{|b\, c|}\right)\, ,
\end{equation}
where $K_\nu(x)$ denotes the modified Bessel-function of order $\nu$.
In this manner, one obtains
\begin{equation}
\label{r49dsecond}
 a^{(1)}_{9,{\rm M};(m\neq 0)}\simeq \frac{2\zeta(3)}{r_{11}^3}+\frac{8\pi}{r^2_{11} \,r_{1}} \sum_{m\ne 0} \sum_{n\ge 1} \left\vert{\frac{m}{n}}\right\vert K_1\left(2\pi  |m| n\, {\frac{r_{1}}{r_{11}}}\right)\,.
\end{equation}
Expressing the two terms \eqref{r49dfirst} and \eqref{r49dsecond}  in
string units, one recovers almost all contributions from the expected
result
\begin{equation}
\label{r49dknown}
a_{9}\simeq \frac{2 \zeta(3)}{g_s^2} +\frac{2\pi^2}{3}\left(1+\frac{1}{\rho^2_{1}}\right)+ \frac{8\pi}{g_s} \sum_{m\ne 0}\sum_{n\ge 1} \left\vert\frac{m}{n}\right \vert K_1\left(2\pi   |m| n\, \frac{\rho_{1}}{g_s}\right)\,,
\end{equation}
except again for the one-loop term $2\pi^2/3$.
The second term in \eqref{r49dsecond} has the correct dependence on
$g_s$ to be interpreted as the contribution from $E\!D0$-brane
instantons wrapping the circle of radius $\rho_1$.
This result can be considered as evidence 
for emergence, arguably even more striking than the previous example, as not only the correct tree-level but also the space-time instantons are recovered from a single Schwinger integral.
The full Schwinger integral including also the longitudinal $M2$-brane, i.e.~the type IIA fundamental string, was evaluated in \cite{deWit:1999ir}, which also reproduced the constant $2\pi^2/3$ term.

\subsubsection*{Emergence of $R^4$-term in 8D}\label{sec:8d}

In 8D, i.e.~$d=8$ and $k=2$, an interesting novelty arises, namely the fact that one has to include also contributions from wrapped transverse $M2$-branes in the Schwinger integral.
As we will see, these are the particle-like states generating the missing $2\pi^2/3$ term. 
The full coefficient of the $R^4$-term in perturbative type IIA theory is \cite{Green:1997as}
\begin{equation}
\label{r48dknown}
\begin{aligned}
a_8 \simeq \frac{2\zeta(3)}{g_s^2}&-\frac{2\pi}{T} \log\left(\rho_2^2\, \left|\eta(iU)\, \eta(iT)\right|^4\right) \\
&+ \frac{8\pi}{\rho_{1}g_s}\!\!\!\!\!\!\!\!\sum_{\substack{m\geq1\\ (n_1,n_2)\neq(0,0)}}\!\!\!\!\!\!\!\!\frac{m}{|n_1+in_2U|}K_1\left(2\pi\frac{\rho_{1}}{g_s}m \,|n_1+in_2U|\right)\,,
\end{aligned}
\end{equation}
with $T=\rho_1\rho_{2}$ and $U=\rho_2/\rho_1$. 
Similarly to the 9D case, the  M-theoretic Schwinger integral with only $D0$-brane contributions gives the tree-level term, the $E\!D0$ instanton corrections, the $U$-dependent part of the one-loop contribution and in principle also the logarithmic contribution $-2\pi\log(\rho_2^2)/T$. More details can be found in \cite{Blumenhagen:2024ydy}.

Besides, one receives a contribution from transverse $M2$-branes carrying KK momentum only along the eleventh direction. 
(Recall that additional KK momentum along $T^2$ would spoil the 1/2 BPS
property.) 
The masses of the particle-like states arising from these wrapped transverse $M2$-branes are
\begin{equation}
M^2=n^2 t_{12}^2+ \frac{m^2}{r_{11}^2}\,,
\end{equation}         
where $t_{12}=r_1 r_{2}$ denotes the area of $T^2$ in M-theory units. 
Thus, the total Schwinger integral also includes a contribution 
\begin{equation}
\label{r48dschwingerD2}
a^{(1)}_{8,{\rm M};M2}\simeq \frac{2\pi}{r_{11} t_{12}}  \sum_{n\ne 0} \sum_{m\in\mathbb Z} \int_0^\infty  \frac{d t}{t}  \;
 e^{ -\pi t \left(  n^2 t_{12}^2 +\frac{m^2}{r_{11}^2}\right)}\, .
\end{equation}
This integral can be regularized following our usual procedure. After performing Poisson resummation with respect to the KK momentum $m$ and applying \eqref{besselrel}, one gets 
\begin{equation}
\label{r48DD2final}
a^{(1)}_{8,{\rm M};M2}\simeq \frac{2\pi}{r_{11} t_{12}} \left( \frac{\pi}{ 3} r_{11} \,t_{12} + 4\!\! \sum_{n_1,n_2\ge 1}  \frac{1}{n_2} e^{-2\pi n_1 n_2 r_{11} t_{12}} \right) = - \frac{2\pi}{T} \log\left( \left|\eta(i T)\right|^4 \right)\,,
\end{equation}
where we used $K_{1/2}(x)=\sqrt{\frac{\pi}{ 2x}}e^{-x}$.
Thus, this $M2$-brane contribution is indeed the missing part so that the complete M-theoretic Schwinger integral, obtained by combining the pure $D0$-brane contribution with \eqref{r48DD2final}, reproduces
\eqref{r48dknown}. Let us notice that in the original work of GGV \cite{Green:1997as}, the term \eqref{r48DD2final} had to be somehow added by hand. In contrast, here it stems from considering wrapped transverse $M2$-branes as required by the physical prescription of integrating out towers of states with typical mass not larger than the species scale.

As explained in \cite{Green:1997as}, the exponential terms in \eqref{r48DD2final} describe type IIA (fundamental) string instantons wrapped on the $T^2$. 
However, let us emphasize once more that in our approach the $M2$-branes are transversely wrapped so that they are not type IIA fundamental string winding modes. 
In addition, $a^{(1)}_{8,{\rm M};M2}$ also contains the constant term $2\pi^2/3$, which was so far missing in 9D and 10D. 
This is completely analogous to the string story,  where our method in 10D was also missing this constant term, while it could be obtained in 9D from integrating out the string winding modes along the compact direction. 
Here, the relevant object is the transverse $M2$-brane which can be particle-like only for $d\le 8$. 
Similar to string theory, upon decompactification of the 8D result, we can  deduce the presence of the same constant term, $2\pi^2/3$, also in 9D and 10D.

\subsubsection*{Comments on emergence in $d\le 7$}

One could now move forward and  consider compactifications on higher dimensional tori. 
The complete evaluation of the Schwinger integral \eqref{fullr4termdef}  becomes increasingly complicated, as there will be additional sectors contributing and the 1/2 BPS conditions become more involved and harder to solve explicitly.
The $d=7$ case is discussed in detail in \cite{Blumenhagen:2024ydy} and confirms the expectation that  the Schwinger integral \eqref{fullr4termdef} produces the full $R^4$-term, including also all instanton corrections, which will now also comprise type IIA Euclidean $E\!D2$-brane instantons.
It would be very interesting to go to $d=5$, where also particle-like $M5$-branes would contribute for the first time. In \cite{Blumenhagen:2024ydy} we only provide partial results, as the full amplitude turns out to be highly complicated.

On the technical level, we have seen that instantons arise via Poisson resummation of certain wrapping numbers and by applying then the relation \eqref{besselrel}. 
In particular, from the argument of the appearing Bessel function we could read off the action of the corresponding instanton.
In this way, we have been recovering the  $E\!D0$ and $E\!F1$ instantons from combinations of $(D0,{\rm KK})$ and $(D2,D0)$ particles upon Poisson resummation of the wrapping number of the second entry. 
Going through all possible such combinations one arrives at table \ref{table_corresp}.
\begin{table}[ht] 
\renewcommand{\arraystretch}{1.5} 
\begin{center} 
\begin{tabular}{|c|c|} 
\hline
Particle states &   Instantons        \\
\hline \hline
$(D0,{\rm KK}_{(k)})$   & $E\!D0_{(k)}$ \\
$(D2_{(ij)},{\rm KK}_{(k)})$   & $E\!D2_{(ijk)}$ \\
$(N\!S5_{(ijklm)},{\rm KK}_{(n)})$   & $E\!N\!S5_{(ijklmn)}$ \\
\hline
$(D2_{(ij)},D0)$   & $E\!F1_{(ij)}$ \\
$(N\!S5_{(ijklm)},D0)$   & $E\!D4_{(ijklm)}$ \\
\hline
$(N\!S5_{(ijklm)},D2_{(lm)})$   & $E\!D2_{(ijk)}$ \\
\hline
\end{tabular}
\caption{Particle (in loop) - instanton correspondence for elementary states.}
\label{table_corresp}
\end{center} 
\end{table}
This constitutes another non-trivial check of the M-theoretic Emergence Proposal.
Even though in the Schwinger integral one only integrates out the perturbative towers of states, all expected instanton actions eventually appear, i.e.~also $E\!F1$, $E\!D4$ instantons which correspond to longitudinal Euclidean $M2$- and $M5$-branes.

\subsection{Emergence of 4D \texorpdfstring{$N=2$}{TEXT} topological couplings}

So far we discussed only a single term in the derivative
expansion of the spacetime low energy effective theory.
The leading order two-derivative terms, like the Einstein-Hilbert term and the kinetic terms for the type IIA $C_1$ and $C_3$ gauge fields, are not 1/2 BPS saturated and currently out of reach with our method.
However, upon compactification on a space that breaks part of the supersymmetry these kinetic terms can become 1/2 BPS saturated as well. 
This happens for compactifications of type IIA string theory on a Calabi-Yau threefold $X$ yielding $N=2$ supergravity in 4D.

Since this is a standard class of models, let us recall some relevant facts just very briefly.   
The resulting  moduli space is locally a product  of  decoupled  hyper- and vector-multiplet moduli spaces. 
The scalar fields in the vector-multiplets are complexified K\"ahler moduli denoted  as $T^i = t^i + i b^i$, with $i=1,\dots, h_{11}(X)$, where $b^i$ are Kalb-Ramond axions, while $t^i$ are real K\"ahler moduli, defining the K\"ahler cone $t^i>0$.
The vector fields are given by the RR $C_3$-form dimensionally reduced on $h_{11}(X)$ homologically two-cycles. There is one more vector field, the graviphoton, residing in the $N=2$ gravity multiplet and given by the type IIA RR $C_1$-form. 
The  kinetic terms for the K\"ahler moduli and the gauge couplings are determined by the holomorphic prepotential ${\cal F}_0(T)$, which is known to only receive perturbative corrections up to one-loop. 
However, for type IIA the prepotential is even completely classical, as the four-dimensional dilaton lies in a hypermultiplet.\footnote{For $K3$ fibrations, the heterotic dual models indeed feature non-vanishing one-loop corrections.}  
In addition, there are non-perturbative corrections from world-sheet instantons, so that the full prepotential enjoys an expansion\footnote{Note that mirror symmetry also fixes the a priori ambiguous quadratic  and linear terms in the prepotential \cite{Hosono:1994ax}.}
\begin{equation}
\label{prepotcompactcy}
{\cal F}_0(T) =-\frac{1}{g_s^2}\bigg[\frac{1}{3!}C_{ijk} T^i T^j T^k+\frac{\zeta(3)}{2}\chi(X) -\!\!\!\sum_{\beta\in H_2(X,\mathbb  Z)} \alpha^{\beta}_0\,\,\,{\rm Li}_3\left(e^{-\beta \cdot T}\right)\bigg]\,,
\end{equation}
where $C_{ijk}$ denote the triple intersection numbers and $\chi(X)$  the Euler characteristic of the Calabi-Yau threefold. 
Moreover, the integers  $\alpha^{\beta}_0$ are the genus zero Gopakumar-Vafa invariants \cite{Gopakumar:1998ii,Gopakumar:1998jq}, which are topological invariants of the Calabi-Yau.
For a recent work relating the Gopakumar-Vafa invariants to the emergent string conjecture see
\cite{Rudelius:2023odg}.

The prepotential is only the first in an infinite series of topological higher derivative couplings of the form $\mathbb{F}_g(T,\ov{T}) R_{+}^2 F_{+}^{2g-2}$, where $R_{+}$ and $F_{+}$ are the self-dual parts of the Riemann tensor and of the graviphoton field strength.  
Up to an additive term independent of the K\"ahler moduli, the coupling splits into a harmonic piece and the so-called holomorphic anomaly
\begin{equation}
\mathbb{F}_g(T,\ov{T})={\rm Re}( \mathcal{F}_{g}(T) ) +  f_g^{anom}(T,\ov T)\, ,
\end{equation}
with $\partial_i\partial_{\bar {\jmath}}  f_g^{anom}(T,\ov T)\ne 0$.
The genus $g$ holomorphic topological string amplitudes $\mathcal{F}_{g}(T)$ are not corrected in type IIA beyond the string $g$-loop level. 

The one-loop topological free energy $\mathbb{F}_1(T,\ov{T})$ is the quantity that was proposed to be a measure for the number of light species, as recalled in section \ref{sec_species}.
Its holomorphic part has an expansion in terms of a linear term and an infinite sum of instanton corrections
\begin{equation}
\label{expandF1}
{\cal F}_1(T)= -\frac{1}{24} c_{2,i}\, T^i - \sum_{\beta\in   H_2(X,\mathbb Z)} \left( \frac{\alpha_0^{\beta}}{12} +\alpha_1^{\beta}\right) {\rm Li}_1(e^{- \beta \cdot T})\,,
\end{equation}
where the  integers  $\alpha^{\beta}_1$ are the genus one Gopakumar-Vafa invariants and $c_{2,i}=\int_X  c_2(T_X)\wedge \omega_i$ denotes the coefficient of the second Chern class of the tangent bundle of $X$, with the K\"ahler form  expanded in a basis of cohomological 2-forms as $J=\sum_{i=1}^{h_{11}}  t_i \,\omega_i$.

\subsubsection*{Emergence of the topological amplitudes}

In the weakly coupled emergent string limit, $g_s\ll 1$, mirror symmetry to type IIB is a very powerful tool to derive the exact expansion \eqref{prepotcompactcy}. 
For a concretely chosen Calabi-Yau, this allows to read off the genus zero Gopakumar-Vafa invariants from the periods upon employing the mirror map.
Like for the $R^4$-term, let us now consider the M-theory limit.
Since the $\mathcal{F}_{g}(T)$ are known to be 1/2 BPS saturated, in this limit one might expect to find a Schwinger-like integral where one integrates out the light towers of states, which are 1/2 BPS bound states of KK momentum along the M-theory circle ($D0$-branes) and transverse $M2$-branes. Indeed, these are the objects that carry charge under the central extensions of the $N=2$ supersymmetry algebra,\footnote{To compare to \eqref{prepotcompactcy} and \eqref{expandF1}, $T$ has to be rescaled by a factor $2\pi$. We have also set $M_s=2\pi$ throughout this calculation.}
\begin{equation}\label{BPS charge}
Z_n(\beta) =\frac{2\pi}{g_s}\left(\beta \cdot T+i n\right) .
\end{equation}
Here $n$ is the number of $D0$-branes and $\beta\in H_2(X,\mathbb Z)$ denotes the homology class wrapped by the $M2$-brane, whose complexified size is $\beta \cdot T$. 
Remarkably, this is precisely the proposal of Gopakumar-Vafa  \cite{Gopakumar:1998ii,Gopakumar:1998jq}, who provided such a Schwinger integral for all of the holomorphic $\mathcal{F}_{g}(T)$. 
For our purposes it is sufficient to concentrate on $\mathcal{F}_{0}(T)$ and  $\mathcal{F}_{1}(T)$, for which Gopakumar-Vafa provided the expressions 
\begin{equation}
\begin{aligned}
  \label{gvschwinger}
  \mathcal{F}_0&= \sum_{\beta}  \alpha^{\beta}_0 \sum_{n\in\mathbb{Z}}\int_{0}^{\infty}\frac{ds}{s^3} \,e^{-sZ_n(\beta)}\,,\\[0.1cm]
\mathcal{F}_1&=-\sum_{\beta} \alpha^{\beta}_1 \sum_{ n\in\mathbb{Z}}\int_{0}^{\infty}\frac{ds}{s}
\,e^{-sZ_n(\beta)}-\frac{1}{12}\sum_{\beta} \alpha^{\beta}_0
\sum_{n\in\mathbb{Z}}\int_{0}^{\infty}\frac{ds}{s} \,e^{-sZ_n(\beta)}\,.
\end{aligned}
\end{equation}
Note that $\mathcal{F}_1$ receives contributions both from genus one
and genus zero curves. In general, the Gopakumar-Vafa invariants $\alpha^{\beta}_g$ count the number of BPS configurations from the transverse $M2$-branes wrapping genus $g$ curves in the class $\beta\in H_2(X)$.

It is important to keep in mind that Gopakumar-Vafa invariants are topological index-like quantities hard to determine from first principles.
Our previous general analysis revealed that there are more light modes in the decompactification limit, which do not carry any central charge appearing in the $N=2$ supersymmetry algebra.
These are discrete KK momenta and transverse $M5$-branes wrapping  4-cycles of the Calabi-Yau.
We observe that the latter would give strings in 4D and upon quantization could result in  contributions to the Gopakumar-Vafa invariants that grow exponentially. Whether this is the right picture remains to be seen.

Evidently, the Schwinger integrals \eqref{gvschwinger} are of the same type as those encountered in the previous section on $R^4$-terms. 
In particular, the integrals for both ${\cal F}_0$ and ${\cal F}_1$ are UV divergent close to $s\simeq 0$, so that we can proceed by regularizing them in the same way as the $R^4$-terms. 
To be concrete, let us focus on two simple examples.

\subsubsection*{The resolved conifold}

First, we look at the non-compact resolved conifold that has only a single $S^2$ of size $T$ on which an $M2$-brane can wrap. 
This can be considered the prototypical example for learning how to evaluate the integrals before eventually taking the infinite sum over all 2-cycles in the homology lattice for a compact Calabi-Yau $X$. 
Here, we only sketch the computation and refer to \cite{Blumenhagen:2023tev} for more details.

Starting with the simpler integral, namely that of $\mathcal{F}_1$, we introduce a UV cut-off and minimally subtract the divergent terms to arrive at
\begin{equation}
\label{f1start}
\mathcal{F}_1^{M2} = \frac{1}{12} \sum_{n \in \mathbb{Z}} \log \left( \frac{ T + in}{\mu}\right)\,,
\end{equation}
with a constant $\mu$ depending on the cut-off and on some numerical factors.
Applying zeta-function regularization for the infinite sum over $n$ and using the identity
\begin{equation}
\label{prodform}
\frac{\sinh(\pi T)}{\pi T}=\prod_{n=1}^\infty \left(1+\frac{T^2}{n^2}\right)
\end{equation}
we can bring this to the form
\begin{equation}
\label{singleF1contri}
 {\cal F}_1^{M2}=\frac{2\pi T}{24} -\frac{1}{12} {\rm Li}_1(e^{-2\pi T})\,.
\end{equation}
We observe that this is identical to the topological free energy computed for the resolved conifold in \cite{Gopakumar:1998ki}, including the linear term.
Formally, the resolved conifold has $c_2=-1$ and the single non-vanishing Gopakumar-Vafa invariant $\alpha^1_0=1$. The sum over only the $D0$-branes gives an ambiguous logarithmic factor $ {\cal F}_1^{D0}\simeq\log(2\pi\mu)$, possibly reflecting the existence of the holomorphic anomaly.

For the holomorphic prepotential, $\mathcal{F}_0$,  one can proceed analogously. After introducing a UV cut-off and minimally subtracting the divergent terms one gets the infinite sum
\begin{equation}
\label{F0-D0-D2-1}
 \mathcal{F}^{M2}_0 = -\frac{2\pi^2}{g_s^2}\sum_{n \in \mathbb{Z}} (T + in)^2 \, \log \left( \frac{T + in}{\mu}\right)\,,
\end{equation} 
with another constant $\mu$, related to the cut-off. 
Again, applying zeta-function regularization and a descendant of the relation \eqref{prodform} obtained after performing two integrations, one finally arrives at
\begin{equation}
\label{prepotfin}
\mathcal{F}_0^{M2}=\frac{1}{g_s^2}  \left[ -\frac{(2\pi T)^3}{12}+{\rm Li}_3(e^{-2\pi T})\right]\,.
\end{equation}
In addition, there is a also a non-trivial contribution from bound states of only $D0$-branes, given by
\begin{equation}
\mathcal{F}_0^{D0}=-\frac{2\pi^2}{g_s^2}\sum_{n\ne
  0}n^2\log\left(\frac{n}{\mu}\right)=
-\frac{1}{g_s^2} \zeta(3)\, .
\end{equation}
The total, unambiguous part of the  prepotential also agrees with \cite{Gopakumar:1998ki}, where the triple intersection number is formally $C=1/2$ and the Euler characteristic is $\chi=2(h_{11}-h_{21})=2$.

One might be worried that this is a too trivial example and that a large portion of the complications for compact Calabi-Yau threefolds is actually absent. These complications involve the sum over the full infinite homology lattice and in particular the a priori unknown index-like Gopakumar-Vafa invariants. 
However, for each individual genus zero curve, the evaluation of the Schwinger integrals will proceed as for the resolved conifold and for both ${\cal F}_0$ and ${\cal F}_1$ we will get a cubic and a linear contribution, respectively. How these contributions do add up to finally give the triple intersection numbers and the second Chern class is far from being  obvious and deserves further investigation.

\subsubsection*{The Enriques Calabi-Yau}

To better illustrate what the challenge of taking the sum over the infinite homology lattice is, let us consider a second simple example, this time on a compact Calabi-Yau.
The  Enriques Calabi-Yau manifold is defined as the free quotient $X=(K3\times T^2)/\mathbb Z_2$, where the $\mathbb Z_2$ acts via a free action on $K3$ and
an inversion $z\to -z$ on the $T^2$.
The free quotient of $K3$ is the Enriques surface ${\cal E}$ with Euler characteristic  $\chi({\cal E})=c_2(T_{\cal E})=12$, leading to a Calabi-Yau with Hodge numbers $(h_{11}(X),h_{21}(X))=(11,11)$. 
This Calabi-Yau is a $K3$ fibration over a base $\mathbb P^1$, with the fibration reducing to ${\cal E}$ over the four $\mathbb Z_2$ fixed points in the base. 
In \cite{Klemm:2005pd,Grimm:2007tm} the one-loop topological free
energy has been computed  using the  duality to the heterotic
string on $K3 \times T^2$ giving
\begin{equation}
\label{Enriques-F1}
{\cal F}_1= -\pi T - 12 \sum_{n\ge 1} {\rm Li}_1(e^{-2\pi n T})+ \ldots\,,
\end{equation}
where we omitted the known contribution depending on the 10 K\"ahler moduli related to the 2-cycles of the Enriques surface.

The question is whether we can reproduce this result by adding up appropriate  $D2/D0$ bound state contributions \eqref{singleF1contri}; we are especially interested in the linear term, $-\pi T$, which is usually obtained upon dimensional reduction but not really from a Schwinger computation. 
Since the action of the $\mathbb Z_2$ orbifold is free on the $K3$, the  relevant 2-cycle is indeed a genus one curve. 
This curve is the $T^2$ sitting at any point of the Enriques surface and which can be multiply wrapped.  
Apparently, its contribution  to the Gopakumar-Vafa invariant $\alpha_1^n$ is just the Euler characteristic of the Enriques surface $\alpha_1^n =\chi({\cal E})=12$, which can be read off from the instanton series in \eqref{Enriques-F1}. 
Taking into account that for the contribution of a genus one curve the prefactor $1/12$ in \eqref{singleF1contri} is replaced by one, the linear term becomes
\begin{equation}
\sum_{n\ge 1} \chi({\cal E}) n\, \pi T = -\pi\, T\,,
\end{equation}
where for the sum over $n$ we employed zeta-function regularization, i.e.~$\zeta(-1)=-1/12$.
Since this expression agrees precisely with the general expansion \eqref{expandF1} for $c_2({\cal E})=12$, by taking the sum over the homology of all individual contributions we have reproduced the known result including in particular the linear term. 
We conclude that the full holomorphic one-loop topological free energy (for the $T$ modulus) emerges just from the Schwinger integral\footnote{The logarithmic ambiguity can be used to include the non-holomorphic term  needed for modular invariance of $F_1(T,\ov T)=-6\log\big( (T+\ov T)|\eta(iT)|^4\big)$.}.

This example reveals also a puzzle. 
In \cite{Klemm:2005pd} it was shown that the prepotential is just 
\begin{equation}
\label{preclass}
{\cal F}_0\simeq-\frac{1}{g_s^2} C_{IJ} T^I T^J T^B, 
\end{equation}
with $C_{IJ}$ denoting the Cartan matrix of $E_8\times \Gamma^{1,1}$ and $T^B$ the K\"ahler modulus of the base $\mathbb P^1$. 
This means  that all genus zero Gopakumar-Vafa invariants are vanishing.

Hence, we face the following question: how can a sum over the individual contributions \eqref{prepotfin} ever lead to a finite cubic term? 
While we have no clear answer yet, considering all the evidence we have collected for emergence, we believe that this puzzle will likely not falsify it but reveal some important subtleties on how emergence is realized  for ${\cal F}_0$.
Recall that the prepotential contains information on kinetic terms. 
Since space-time is expected to emerge together with these terms\footnote{We thank Ivano Basile and Eran Palti for comments on this point.}, it is conceivable  that one has to go beyond this quasi-geometric approach, with (BPS) branes wrapping cycles, to reliably compute  ${\cal F}_0$.

Perhaps this is also related to the more radical proposal made in \cite{Hattab:2023moj}, where the initial real Schwinger integral for ${\cal F}_0$ was deformed in two ways. 
First, the integral was taken over a contour in the complex plane and the integrand was changed from the simple exponential form, $\exp(-s Z_n(\beta))$, so that in the end one arrives at a contour representation of ${\cal F}_0$ familiar from period computations. 
It was shown that this yields the full expansion of the prepotential as derived from the period computation in the weakly coupled emergent string limit, $g_s\ll 1$. This calculation was supported and refined by relating the UV degrees of freedom that are being integrated out to Fermi gas degrees of freedom \cite{Hattab:2024thi}. The relation between this and our approach remains to be investigated.
We have now reached our present level of understanding of the subject and hence refer the reader to future research.

\section{Final comments on emergence}

In the previous sections, we reviewed our current indications for the M-theoretic Emergence Proposal. 
These are based mainly on the evaluation of certain 1/2 BPS saturated couplings that were also subject to perturbative non-renormalization theorems beyond the one-loop level. 
In these cases it was possible to recover the full couplings in a small $g_s$ expansion from just a single one-loop Schwinger integral in M-theory, where one integrates out solely the light, perturbative towers of particle-like states with masses not larger than the species scale, i.e.~the 11D Planck scale.
From where we stand, let us reflect on two natural next questions.

First let us recall that, in string theory, by implementing the 1/2 BPS condition via a Lagrange multiplier and using modular invariance, one could define a complex version of the Schwinger integral that also gives the correct constant term, $2\pi^2/3$, in 10D. 
The non-trivial question is whether one can generalize this to M-theory and as such provide a higher-dimensional definition of the Schwinger integrals that upon evaluation also yields the correct result in 10D and 9D.
The essential difference to the string-theoretic case is that, in M-theory, the number of 1/2 BPS conditions increases with the number $k$ of compactified directions. 
As we have seen, the latter transform in the representation
$\lambda_{E_k}$ of the group $E_{k(k)}$ so that after imposing them via Lagrange multipliers, we arrive at an integral over ${\rm dim}(\lambda_{E_k})+1$  dimensions. 
The question is whether this can be interpreted as the moduli space ${\cal M}_k$ of something like an $M2$-$M5$ world-sheet.
At the moment, this is not obvious (to us) at all and it might not even be
the right way to proceed. 
Indeed, the BFSS matrix model has taught us that the quantization of the $M2$-brane should rather be thought of as a second quantized theory and not as a first quantized world-volume theory, like for the string.

Second, the Schwinger integral approach  has been performed in this still sort of geometric manner thanks to the 1/2 BPS nature of the involved states.
The ultimate  question is whether emergence is just a special aspect of such a simplified set-up or whether it is a general property of M-theory. 
In the latter case, all terms in the low energy effective action, including  the 10D tree-level kinetic terms, have to arise from quantum effects.
However, the high degree of supersymmetry forbids such couplings to be generated via loops so that we cannot expect to find a simple Schwinger integral that gives these second derivative couplings right away.

Nevertheless, to conclude that this already spoils the M-theoretic Emergence Proposal might be premature for the following reasons.
First, for non-BPS saturated couplings, like the kinetic terms in 10D with maximal supersymmetry, all the excitations of M-theory will matter and contribute. 
The only candidate we have at present to deal with this situation is the BFSS matrix model. 
Second, recall that also in string theory one is computing the low-energy effective action only indirectly. Indeed, one has to compute the appropriate on-shell scattering amplitudes that allow to gain information about the underlying EFT.  
This is done by comparing  part of the string amplitude with the corresponding amplitude computed in the expected or proposed EFT. 
Therefore, to see the emergence of the 2-derivative kinetic terms one has to compute the appropriate object. 
In the context of the BFSS matrix model this problem has already been approached.
Here, we are not reviewing all these efforts, but we just recall that in such matrix model two gravitons (i.e.~two $D0$-branes) do not interact classically. Instead,  the leading order interaction is generated at the one-loop  level (in matrix theory) leading to an effective potential
\begin{equation}
\label{gravipot}
V=-\frac{15}{16} \frac{v^4}{r^7}+\ldots\,,
\end{equation}
where $v$ denotes the relative velocity of the two gravitons.
Note that a non-vanishing $v$ breaks supersymmetry.
This potential  is precisely the leading order classical interaction in the supergravity theory. Due to supersymmetry, the latter is vanishing for $v=0$, i.e.~the forces due to graviton and $p$-form exchange do cancel.
Only upon breaking supersymmetry one gets a non-trivial potential but this is generated at one-loop in the M(atrix) theory.
We think that this is very much in accordance with the Emergence Proposal and also features the expected correlated emergence of the kinetic terms and space itself, here from non-commutative matrix degrees of freedom. 
As a word of caution, it is not a priori clear that the BFSS matrix model is the theory that emerges in the coscaled  infinite distance limit we are taking. 
This deserves further study.

Connecting to the concept of emergence from the very beginning of this article, we note that the behavior \eqref{gravipot} can already be derived from the familiar annulus diagram \cite{Douglas:1996yp} for an open string connecting two $D0$-branes with relative velocity $v$ and distance $r$.  
Hence, we are tempted to speculate that the M-theoretic Emergence Proposal might be closely related to the celebrated loop-channel tree-channel equivalence for the $D0$-branes (and $D2$-branes), which happen to be the lightest states in the $M_* R_{11}\gg 1$ region of M-theory. In this respect,  the transverse $M5$-brane, i.e.~the type IIA $NS5$-brane, is of a different type, pointing to the main open issue  of  the BFSS matrix model.

\paragraph{Acknowledgments.}
We thank Carlo Angelantonj, Ivano Basile, \'Alvaro Herr\'aez, Elias Kiritsis, Wolfgang Lerche, Dieter L\"ust, Eran Palti, Andreas Schachner, Timo Weigand and Max Wiesner for useful discussions. The work of R.B.~and A.G.~is funded by the Deutsche Forschungsgemeinschaft (DFG, German Research Foundation) under Germany’s Excellence Strategy – EXC-2094 – 390783311. R.B.~and N.C.~thank the hospitality of the  Corfu Summer Institute 2023, where part of the research reviewed here has been performed.

\end{document}